\documentclass[seceq]{jpsj2} 
\title{Mott Plateau in Trapped Bose-Fermi Mixture}

\author{Akiko Masaki, Sota Tsukada and Hiroyuki Mori}

\inst{Department of Physics, Tokyo Metropolitan University,\\
Minamiohsawa 1-1, Hachioji-shi, Tokyo 192-0397, JAPAN}

\abst{Boson-Fermion mixture confined on an optical lattice is numerically studied. Mott-plateau like behavior is observed under certain conditions in the density profile of the bosons and fermions as one sees in bosonic atoms in a trap potential. In the Mott plateau area, the total number of the bosons and fermions at each site is kept to an integer value but both local density of the bosons and that of the fermions take some intermediate values. Unlike bosonic or fermionic Mott insulators, the bosons and fermions can move even in the Mott plateau area by exchanging their positions. With Monte Carlo simulations we demonstrate the formation of the Mott plateau, examine the property of the bosons and fermions by measuring the densities and local density fluctuations, and give discussion on this rather new state.}

\kword{ultra-cold atomic gases, Mott transition, optical lattice, fermion-boson mixture}

\begin{document}
\maketitle

\section{Introduction} 
Remarkable progress of laser technologies in the last few decades has realized various fundamentally-new quantum states of ultracold atoms confined in a certain space. The interatomic interactions play sometimes significant roles there. One of the drastic demonstrations of such roles can be found in Mott insulating phase of atoms on optical lattices. Mott insulating phase appears when the criteria, i.e. sufficiently strong interatomic interactions and commensurable ratio of atoms and lattice sites, are met. However, unlike those in a flat potential, which have been well investigated, atoms in a confinement potential exhibit quite different features\cite{campbell,batrouni,wessel}.

Unconfined atoms interacting on a lattice could form Mott insulating state when their average number density per site coincides with an integer value. Confined atoms, on the other hand, distribute nonuniformly over the lattice in the presence of a trap potential, with the density profile whose peak is located in the deepest position of the potential. Therefore the commensurability criteria cannot be satisfied globally but only locally. Assume that atoms are put in an isotropic confinement potential plus periodic optical potential. The density of the atoms would be peaked at the center of the confinement potential and decreases as a function of the radius from the center. At some point of the radial direction, the density may reach an integer value, locally satisfying the criteria of the commensurability. Around this area, the atoms would form local Mott state with sufficiently large repulsive interactions. If we look at the density profile in the radial direction in this case, there should be a local plateau, called Mott plateau, at and around the commensurate-density point. This produces a shell structure of the Mott phase of the confined atoms.

The studies on the atomic Mott phases have been conducted mostly for bosonic atoms, particularly because of the relative ease of the experiments. However, the current technology even realizes mixture systems of bosons and fermions in a continuum space \cite{Schreck,Hadzibabic,Roati,Ferrari} and on a lattice\cite{Modugno}, and so we may expect Mott states to form in the mixture systems as well under certain conditions. Actually the possibility of untrapped boson-fermion mixture to undergo Mott transition was pointed out in numerical calculations\cite{takeuchi}, which indicated that the Mott transition occurred with sufficiently large boson-boson interactions although the bosons and fermions were mixed or demixed depending on the strength of boson-fermion interactions. It is hence natural to consider that the shell structure of the Mott phase should be observed in the confined boson-fermion mixtures, and this is the main objective of the present study.

To accomplish this goal, we performed quantum Monte Carlo simulations of bosons and fermions on a lattice in a harmonic confinement potential. By tuning interactions, particle number, potential curvature, we realized Mott plateau in the boson-fermion mixture, where the total local density of the bosons and fermions was fixed to an integer value, but either local density of the bosons or that of the fermions was below the integer value. To study the local property of the mixture system, we also observed local density fluctuations.

The paper is organized as follows. Section 2 is devoted to the introduction of the model that we used in the simulation. Brief explanation on the local compressibility is also given in the section. In Section 3 we show our numerical result and Section 4 is devoted to the discussion on the Mott plateau. Summary and discussions are given in Section 5.
\section{Model}
We performed world-line quantum Monte Carlo simulation of confined bosons and fermions on an optical lattice\cite{hirsch,batrouni1,batrouni2}. For simplicity the particles were placed on a one-dimensional lattice of $L$ sites. The Hamiltonian that we employed for the simulation was
\begin{eqnarray}
H&=&-t_b\sum_ib^\dagger_ib_{i+1}+h.c.-t_f\sum_if^\dagger_if_{i+1}+h.c.\nonumber\\
&&+\frac{U_{bb}}{2}\sum_in_{bi}(n_{bi}-1)+U_{bf}\sum_in_{bi}n_{fi}\nonumber\\
&&+V_c\sum_i(i-L/2)^2(n_{bi}+n_{fi}).
\end{eqnarray}
Here $b (b^\dagger )$ and $n_b$ are boson annihilation (creation) operator and boson number operator respectively, and $f (f^\dagger )$ and $n_f$ are fermion annihilation (creation) operator and fermion number operator respectively. $U_{bb}$ denotes the boson-boson interaction and $U_{bf}$ the fermion-boson interaction. $V_c$ is the curvature of the quadratic confinement potential. Fermion-fermion interactions are ignored here. Assuming all the interatomic interactions were repulsive, we observed density profile and local compressibility of the boson-fermion mixtures to find local Mott states. We define the local compressibility $\kappa_i$ as local density fluctuation at site $i$, expressed by
\begin{eqnarray}
\kappa_{bi}&=&\beta (\langle n_{bi}^2\rangle -\langle n_{bi}\rangle^2),\nonumber\\
\kappa_{fi}&=&\beta (\langle n_{fi}^2\rangle -\langle n_{fi}\rangle^2),\\
\kappa_{bfi}&=&\beta (\langle (n_{bi}+n_{fi})^2\rangle -\langle (n_{bi}+n_{fi})\rangle^2).\nonumber
\end{eqnarray}
Here $\kappa_{bi}$ is the fluctuation of the density of the bosons at the site $i$, $\kappa_{fi}$ is that of the fermions, and $\kappa_{bfi}$ represents the fluctuation of the total density of the bosons and fermions. 
As pointed out previously\cite{wessel}, the local compressibility defined by $\partial\langle N\rangle /\partial\mu_i$ where $\mu_i$ is the local chemical potential at the site $i$ is more sensitive to distinguish the local Mott state than that defined by the density fluctuation used in the above. Since the former definition requires fluctuation of the total number of atoms in the simulation, which is not feasible with our technique, we employed the latter definition for the local compressibility. The local compressibility of the latter definition does not exactly reach zero even in the local Mott state, but it is still sufficiently sensitive to detect the state.

All the simulation results shown below were obtained for 35 bosons and 30 fermions on 80 sites. Temperature and Trotter decomposition number were fixed to $T=0.08$ and $L_\tau =100$, respectively, and the hopping energies of the bosons and fermions were set to $t_b=t_f=1$ as energy unit.
\section{Simulation result}
\begin{figure}[tb]
\begin{center}
\resizebox{80mm}{!}{\includegraphics{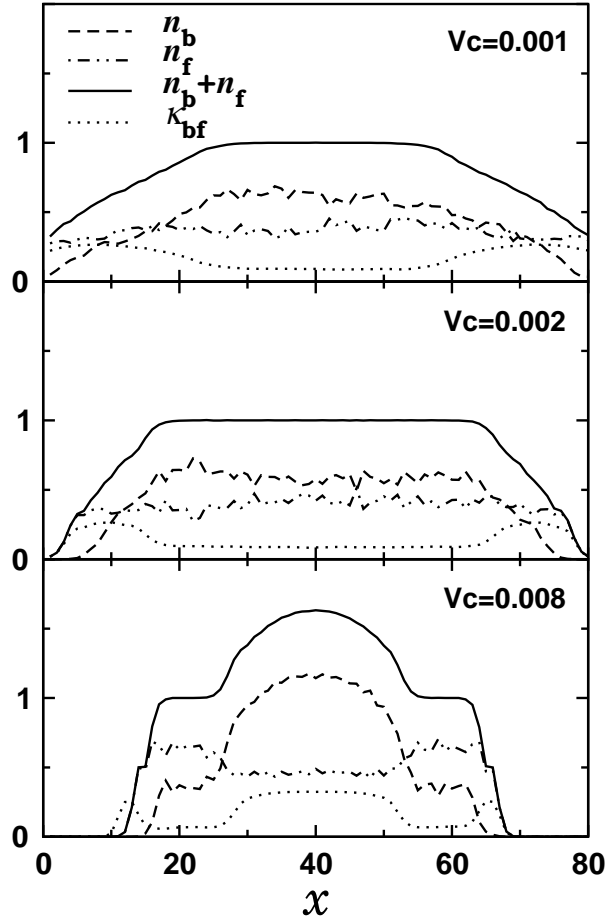}}
\end{center}
\caption{For various $V_c$, the density profiles of the boson ($n_b$) and fermion ($n_f$) are shown with the total density ($n_b+n_f$). The local compressibility $\kappa_{bf}$ of the bose-fermi mixture, defined in the text, is also illustrated. There is a plateau area in the total density profile with small compressibility, indicating that the area is in the Mott state.}
\label{fig1}
\end{figure}

We have many controllable parameters in the present system such as the interactions, particle number, and potential shape. We thus can obtain numerous results by changing the parameter set, although we here focus on typical cases where only the total local density, namely the sum of the boson density and fermion density, exhibits Mott plateau, and the boson density or fermion density does not have the plateau. We fixed the interaction parameters to $U_{bb}=U_{bf}=8$, with which the bosons and fermions were expected to be mixed if not confined, according to the previous calculation\cite{takeuchi}. Figure \ref{fig1} shows the density profiles of the bosons and fermions, $n_b(x)$, $n_f(x)$, and $n_b(x)+n_f(x)$. $x=40$ is the center of the trap potential. The solid line, presenting the total density $n_b+n_f$, has Mott plateau where the density is fixed to 1. The Mott-like feature is also demonstrated by the local compressibility $\kappa_{bf}$, which almost diminishes in the Mott plateau area. As pointed out in the above, the local compressibility does not go to zero strictly but clearly shows local incompressible areas. As we go out from the Mott plateau area, the total density per site no longer takes the integer value and at the same time the compressibility is enhanced, indicating the existence of the superfluidity of the bosons and the metallic state of the fermions.

Figure 1 also presents the $V_c$ dependence of the density profiles. When the curvature $Vc$ of the confinement potential increases, all the particles are forced more strongly to move toward in the center area and participate in the formation of wider Mott plateau. When $Vc$ exceeds a critical value, which is dependent of the interactions and the number of particles, a bump appears in the center of the density profile to save the confinement potential energy although paying the extra cost for the interaction energy, as can be seen in the bottom graph of the figure. Since more than one boson can occupy a single site simultaneously if paying extra cost due to the interaction energy, and since the double occupancy is strictly prohibited for the fermions, the bosons have stronger tendency of accumulating in the center of the potential. This accumulation of the bosons in the center avoids the distribution of the fermions in the same area by the boson-fermion interaction $U_{bf}$. Therefore many bosons come around the center area to save the potential energy, but the fermions cannot save the potential energy by doing the same thing because of the prohibition of the double occupancy. So the fermions choose to save the boson-fermion interaction energy by escaping from the boson-accumulated area. This behavior can be clearly seen in the bottom graph of the figure, where the boson density has a peak at the center and the fermion density is relatively low in the same area but increases in the surrounding area.

What will happen if we change the interaction parameters $U_{bb}$ and $U_{fb}$? We skipped showing the simulation results obtained for other values of the interaction parameters, as they are not qualitatively different from those shown here. The $U_{bb}$ or $U_{bf}$ dependency of the result is rather simple. With the increase of the interaction between the bosons, $U_{bb}$, the density profile of the bosons in the figure becomes broader for them to keep away from each other. This has negative effect to the formation of the Mott plateau. When $U_{bb}$ is decreased, on the other hand, more and more bosons can gather around the center of the potential, forming a bump as seen in the bottom graph of the figure. The boson-fermion interaction $U_{bf}$ also works against the Mott plateau development, since the bosons gathering around the center of the potential would try to avoid the fermions to save the interaction energy. With sufficiently large $U_{bf}$, demixing of the bosons and fermions occurs, as observed in the unconfined boson-fermion mixtures\cite{takeuchi}. Therefore the formation of relatively large Mott plateau is realized with moderate values of $U_{bb}$ and $U_{bf}$, which are dependent of $V_c$.

\section{Mott plateau}

In bosonic systems, the coexistence of superfluid and Mott regions in trapped inhomogeneous systems was already indicated in experiments\cite{campbell}  and theoretical calculations\cite{batrouni,wessel} . Here we demonstrated similar coexistence in the bose-fermi systems, although the definition of the "Mott region" is slightly different. The Mott region in the present system is the area where only the total density of the bosons and fermions is fixed to a certain integer value. The density fluctuation of individual species is basically allowed there if the density fluctuation of the bosons is compensated by that of the fermions to fix the total density. This feature can only be realized in the mixture systems. As pointed out in the thorough study of bosonic systems\cite{wessel}, the Mott plateau emerges not due to a quantum phase transition but in a crossover process. Namely the Mott phase's volume fraction grows while the volume fraction of the superfluid decreases in the crossover. This is quite different from unconfined systems.

Unlike bosonic or fermionic Mott insulators where the particles cannot move and are frozen on each site, the bosons and fermions in the Mott plateau area in the mixture system can move. All the sites in this region are occupied by either a boson or fermion, but this does not necessarily mean that all the particles are frozen in their positions forming an "insulating" area. When a boson and a fermion are located next to each other, they can exchange their positions by going through a virtual doubly-occupied state. The transfer energy of such particle exchange is therefore approximately given by $t\sim t_ft_b/U_{bf}$. With this particle-exchange energy, the bosons and fermions are capable of moving in the Mott plateau area, which is quite different feature from the insulating behavior of bosonic or fermionic Mott state.

\begin{figure}[tb]
\begin{center}
\resizebox{80mm}{!}{\includegraphics{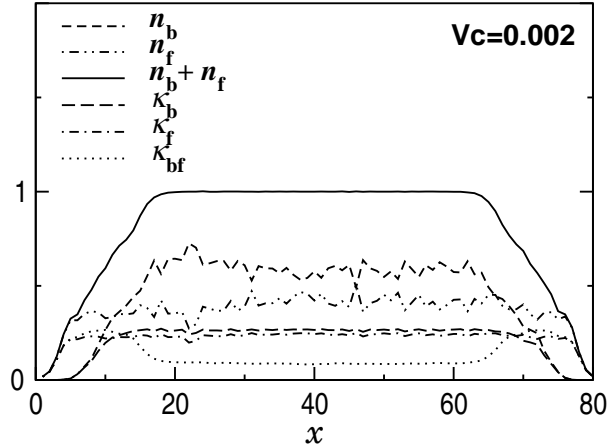}}
\end{center}
\caption{The second graph of the Figure 1 is shown again with additional lines of the local compressibility of the bosons and that of the fermions.}
\label{fig2}
\end{figure}

The "locally compressible" state of the bosons and fermions in the Mott plateau can be observed in the local density fluctuation measurement. The density fluctuation of the bosons $\kappa_b$ and that of the fermions $\kappa_f$ are presented in Fig. \ref{fig2}. We see from the figure that the density fluctuation of either particles takes some finite value even in the Mott plateau area, indicating that the bosons and fermions are not "rigid".
\section{Summary and discussion}
We performed Monte Carlo simulations of the boson-fermion mixture systems on a periodic optical lattice placed in a harmonic potential. The density profile exhibited the coexistence of local Mott state and superfluid (fermi liquid) with appropriate set of parameters. In the local Mott state, the total density of the bosons and fermions took an integer value, while the density of individual species did not. 

To get a more simple view on this local Mott state, let us consider an extreme case. Suppose that we have a hardcore boson system with the number of the bosons being less than the number of the lattice sites. This system exhibits superfluidity. Next we place spinless fermion in each vacant site and prohibit the bosons and fermions to come to the same site. Namely every site is occupied by a single boson or single fermion and there is no empty or doubly-occupied site. The bosons and fermions locating next to each other are however allowed to exchange their positions with the transfer energy $t$. The bosons and fermions can thus move around keeping the number of particle at each site to 1. This does not straightforwardly mean that the bosons condensate at a low temperature. To have a long-range order of the phase of single boson field operator, the number of the bosons has to fluctuate, which is not allowed because of the local constraint on each site. This simple view reminds us a technique called slave-boson (or slave-fermion) method, which has been extensively used for the analysis of strongly correlated electron systems such as $t$-$J$ model. The analysis with the slave-boson (slave-fermion) method has a rather long history and revealed interesting, sometimes peculiar or even unphysical, properties depending on the approximation used. We therefore need careful, detailed analysis on the property of the present system. Analytical work in this direction is in progress.
\section*{Acknowledgment}
The authors would like to thank Professors R. Shiina and T. Hotta for fruitful discussions and Professor N. Kawakami for valuable comments.
%
%


\begin{thebibliography}{99} 
\bibitem{campbell} G. K. Campbell, J. Mun, M. Boyd, P. Medley, A. E. Leanhardt, L. G. Marcassa, D. E. Pitchard, W. Ketterle: Science \textbf{313} (2006) 649.
\bibitem{batrouni} G. G. Batrouni, V. Rousseau, R. T. Scalettar, M. Rigol, A. Muramatsu, P. J. H. Denteneer, and M. Troyer: Phys. Rev. Lett. \textbf{89} (2002) 117203.
\bibitem{wessel} S. Wessel, F. Alet, M. Troyer, And G. G. Batrouni: Phys. Rev. A\textbf{70} (2004) 053615.
\bibitem{Schreck}F. Schreck, L. Khaykovich, K. L. Corwin, G. Ferrari, T. Bourdel, J. Cubizolles, and C. Salomon: Phys. Rev. Lett. {\bf 87} (2001) 080403.
\bibitem{Hadzibabic}Z. Hadzibabic, C. A. Stan, K. Dieckmann, S. Gupta, M. W. Zwierlein, A. G\"{o}rlitz, and W. Ketterle: Phys. Rev. Lett. {\bf 88} (2002) 160401.
\bibitem{Roati}G. Roati, F. Riboli, G. Modugno, and M. Inguscio: Phys. Rev. Lett. {\bf 89} (2002) 150403. 
\bibitem{Ferrari}G. Ferrari, M. Inguscio, W. Jastrzebski, G. Modugno, G. Roati, and A. Simoni: Phys. Rev. Lett. {\bf 89} (2002) 053202.
\bibitem{Modugno}G. Modugno, F. Ferlaino, R. Heidemann, G. Roati, and M. Inguscio: Phys. Rev. A {\bf 68} (2003) 011601. 
\bibitem{takeuchi} Y. Takeuchi and H. Mori: Int. J. Mod. Phys. B\textbf{20}  (2007) 617.
\bibitem{hirsch}J. E. Hirsh, R. L. Sugar, D. J. Scalapino, and R. Blakenbecler: Phys. Rev. B\textbf{26} (1992) 9051.
\bibitem{batrouni1}G. G. Batrouni, R. T. Scalettar, and G. T. Zimanyi: Phys. Rev. Lett. \textbf{65} (1990) 1765.
\bibitem{batrouni2}G. G. Batrouni, R. T. Scalettar: Phys. Rev. B\textbf{46} (1992) 9051.
\end{thebibliography}
\end{document}